\documentclass[aps, preprint]{revtex4-1}

\usepackage{amssymb}

\begin{document}
\title{Primordial Fluctuations from Inflation with a Triad of Background Gauge Fields}
\date{\today }
\author{Kei Yamamoto}
\email{K.Yamamoto@damtp.cam.ac.uk}
\affiliation{DAMTP, Centre for Mathematical Sciences, University of Cambridge, \\
Wilberforce Road, CB3 0WA, United Kingdom}

\begin{abstract}
We study the linear perturbation of the recently proposed model of inflation where a uniform 
gauge-kinetic coupling of the inflaton to multiple vector fields breaks the cosmic no-hair 
conjecture while maintaining the isotropy. We derive the general quadratic action for the 
perturbation and calculate the power spectra of scalar and tensor modes at the end of 
inflation by in-in formalism. It is shown that the model predicts slightly red spectra and the 
tensor-to-scalar ratio tends to be suppressed. The comparison with the data from WMAP 
7-year does not impose strong constraints on the parameters and both weak- and strong- 
gauge-field regimes are consistent with the current observations. 
\end{abstract}

\pacs{98.80.Cq, 98.80.Jk}

\keywords{vector inflation, cosmological perturbation, gauge-kinetic coupling}

\maketitle

\section{Introduction}
Over the past decade, the prospect of unveiling minute details of density fluctuations through 
observations of the Cosmic Microwave Background (CMB) and large scale structure has driven a growing 
interest in inflationary scenarios beyond the single-field slow-roll model. Despite its simplicity and 
the remarkable success in matching the observed density power spectrum, it is hard to believe 
that a single scalar field is entirely responsible for the dynamics of the early universe. Along with multi-scalar 
models such as hybrid inflation \cite{hybrid}, attempts have been made to incorporate vector fields \cite{Ford,Koivisto,Mukhanov}, which have 
proven to be largely unsuccessful due to various instabilities in realizing accelerated expansion \cite{Contaldi1, Mota, Contaldi2, Golovnev, Uzan}. 
Recently, a viable model of vector inflation has been proposed by Soda and his collaborators \cite{Soda1}
where the gauge-kinetic coupling with inflaton, originating from supergravity, maintains the 
amplitude of the vector and results in non-trivial signatures in the primordial fluctuations, e.g. 
statistical anisotropy and scalar-tensor correlations \cite{Soda2}. This "inflation with vector-hair" has been
subsequently scrutinized \cite{Soda3, Soda4, Moniz, Murata, Do2, Hassan} and its stability has been widely established \cite{Dimopoulos, Hervik,Do}. Moreover, it was
found that the inclusion of additional vector fields tends to reduce anisotropy in the background
through a generic dynamical mechanism \cite{minhair}. In the case of uniform gauge-kinetic coupling for three 
or more vectors, the trajectories converge to a universal isotropic attractor with non-vanishing
vector energy density, whose amplitude is determined by the strength of the coupling. 
Very similar dynamical behaviors have been observed in the models employing non-Abelian 
gauge fields \cite{Iran, Wyman, Peter, Jabbari}, which indicates that the tendency towards 
isotropy is generic in the inflation with multiple vectorial degrees of freedom. 
This special case is indistinguishable from the single-field slow-roll model at the level of 
classical background dynamics as long as it provides a sufficient e-folding number. As such,
it is necessary to investigate its linear perturbation in order to impose observational constraints
and decide its viability, which is the aim of this article.  

Incidentally, these gauge-kinetic models can be viewed as the classical counterpart for the 
particle production effects in preheating scenarios. In this context, couplings between the inflaton 
and gauge fields have 
often been discussed in the attempts of generating primordial magnetic fields \cite{Ratra, Demozzi}, gravitational waves \cite{Sorbo}, 
and non-Gaussianity \cite{Peloso}. The crucial conceptual leap of the inflationary models considered in this
article is the breach of 
the cosmic-no-hair conjecture whereby typical preheating scenarios assume vanishing classical
background values of the fields excited by the inflaton, which forces the analysis to go beyond 
linear order and makes it complicated. The presence of vector-hair in inflation illustrates that 
the non-vanishing background energy density of (potentially anisotropic) auxiliary fields does not
necessarily mean a break down of the inflationary regime and the accelerated expansion may
continue without wiping out classical "hair," being the attractor solution in the phase space 
at the same time. Therefore, it is interesting to ask whether the existence of the background
gauge fields can produce any of the features mentioned above within the well-established 
linear perturbation around the isotropic Friedmann-Lema\^itre-Robertson-Walker space-time 
and quantum field theory  in the quasi-de-Sitter background, and how large is the amplitude if any. 

In the present article, we calculate the spectra of the scalar curvature in uniform density slicing and the gravitational waves generated through the interaction of the perturbative variables with the 
background triad of the gauge fields. The result obtained is very much analogous
to what was found for the anisotropic cases \cite{Dulaney, Himmetoglu, Soda2}. We find that the scalar power spectrum acquires 
a double logarithmic scale dependence with the magnitude of the correction being given by the 
fractional energy density of the background gauge fields with respect to the scalar kinetic energy. The effect 
is not slow-roll suppressed due to the steepness of the gauge-kinetic function that is needed 
to maintain the background gauge fields in the accelerated expansion. 
While the tensor mode receives similar corrections, 
the magnitude is smaller than that for the scalar mode by a factor of slow-roll parameter, which was also 
seen in the anisotropic cases. However, since the effect is completely isotropic, the restrictions 
coming from the observational data are much weaker than the anisotropic single-gauge-field models. 
Although the correction terms involve e-folding numbers, which resulted in strong constraints 
on the energy density of the gauge field in the anisotropic models, the large face-vale of the 
e-folding numbers can only affect the overall amplitude of the power spectra in this isotropic 
setup. Since the amplitudes of the quantum fluctuations are always normalized by the energy
scale of inflation, the potentially large corrections to these amplitudes can be absorbed into
this normalization, whence do not immediately pose a problem.

The paper is organized as follows. In section II, we present the background equations and 
introduce the slow-roll parameters to characterize the inflationary dynamics. Section III explains
the structure of the second order action for the gauge fields, which lead to an extended notion
of scalar-vector-tensor decomposition. In section IV, the power spectrum of the scalar mode 
is computed by employing in-in formalism on the de-Sitter space-time. In section V, a similar 
calculation is performed for the tensor mode and observables such as spectral tilt and 
tensor-to-scalar ratio are obtained. Section VI summarizes the results and discusses the outlook.

\section{The Isotropic Background Dynamics of the Inflation with Gauge-kinetic Coupling}

We consider the universe described by the following action;
\begin{equation}
S = \int dx^4 \sqrt{-^4 \! g}\left( \frac{M^2_{\rm pl}}{2}R -\frac{1}{2}\partial _{\mu }\varphi \partial ^{\mu }\varphi - V(\varphi ) -\frac{f(\varphi )^2}{4}F^{(m)}_{\ \ \mu \nu }F^{(m)\mu \nu } \right) . \label{eq:action}
\end{equation}
The scalar field $\varphi $ acts as the inflaton and interacts with the gauge fields through the 
gauge-kinetic function $f(\varphi )$. Although we assume for simplicity that 
$F^{(m)}_{\mu \nu }, (m = 1,2,3)$ are three copies of Abelian gauge field
\begin{equation}
F^{(m)}_{\ \ \mu \nu } = \left( d A^{(m)} \right) _{\mu \nu},
\end{equation}
there is strong evidence for the fact that the dynamics of an $SU(2)$ gauge field with a gauge-kinetic coupling 
in an inflationary regime is well described by the action (\ref{eq:action}) \cite{Murata}. The generalization 
for $m \geq 4$ is also straightforward. We adopt the ADM formalism and follow the notation of 
the ref. \cite{ADM} to write the metric as
\begin{equation}
^4 \! g_{\mu \nu } = \left(   \begin{array}{cc}
    -N^2 + N_k N^k  & N_j \\ 
    N_i  & g_{ij} \\ 
  \end{array} \right)
\end{equation}
where
\begin{equation}
g^{ik}g_{kj} = \delta ^i _{\ j}, \ \ \ \ \ N^i = g^{ij}N_j .
\end{equation}
It is well known that in terms of the normalized extrinsic curvature
\begin{equation}
E_{ij} = -\frac{1}{2}\left( \dot{g}_{ij}-2N_{(i|j)} \right)
\end{equation}
and the intrinsic scalar curvature of the constant time slice
\begin{equation}
^3 \! R = \left( g_{ij,kl}+g_{mn}\Gamma ^m_{\ ij}\Gamma ^n_{\ kl}\right) \left( g^{ik}g^{jl}-g^{ij}g^{kl} \right) ,
\end{equation}
the gravitational part of the Lagrangian is written as
\begin{equation}
\frac{1}{M_{\rm pl}^2}\mathcal{L}_g = \frac{\sqrt{g}}{2N}\left( E_{ij}E^{ij} -E^2 \right) + \frac{1}{2}N\sqrt{g}\ ^3 \! R .
\end{equation}
We introduce the electric fields by
\begin{equation}
E_i^{(m)} = F^{(m)}_{\ \ 0i}
\end{equation}
for the $1+3$ decomposition and write the gauge-field Lagrangian in the form
\begin{equation}
\mathcal{L}_M = \frac{\sqrt{g}}{2N}f^2 g^{ik}\left( E^{(m)}_{ i}+ F^{(m)}_{\ \ ij}N^j \right) \left( E^{(m)}_{k} + F^{(m)}_{\ \ kl}N^l \right) - \frac{1}{4}N\sqrt{g}f^2 g^{ik}g^{jl}F^{(m)}_{\ \ ij}F^{(m)}_{\ \ kl} . \label{eq:maxwell}
\end{equation}
As usual, the scalar Lagrangian is given by
\begin{equation}
\mathcal{L}_{\varphi } = \frac{\sqrt{g}}{N}\left( \frac{1}{2}\dot{\varphi }^2 - \dot{\varphi }\varphi _{,i}N^i + ( \varphi _{,i}N^i )^2 \right) -N\sqrt{g} \left( \frac{1}{2}g^{ij}\varphi _{,i}\varphi _{,j}+V(\varphi ) \right) .
\end{equation}

For the background, we use the ansatz
\begin{equation}
N = \mathcal{N}(t),  \ \ \ N_i = 0, \ \ \ g_{ij} = a(t)^2 \delta _{ij} ,  \ \ \ \varphi = \bar{\varphi }(t), \ \ \ A^{(m)}_0 = 0,  \ \ \ A^{(m)}_{i} = A(t)\delta ^m_{\ i} .
\end{equation}
 The resulting equations of motion read
 \begin{eqnarray}
&& M^2_{\rm pl}\left( \frac{\dot{a}^2}{a^2}+2\frac{\ddot{a}}{a}-2\frac{\dot{a}}{a}\frac{\dot{\mathcal{N}}}{\mathcal{N}}\right) +\frac{1}{2}\dot{\bar{\varphi }}^2 -\mathcal{N}^2 V +\frac{f^2}{2a^2}\dot{A}^2 =0, \\
&& 3M^2_{\rm pl}\frac{\dot{a}^2}{a^2}-\frac{1}{2}\dot{\bar{\varphi }}^2-\mathcal{N}^2 V -\frac{3f^2}{2a^2}\dot{A}^2 =0, \\
&& \ddot{\bar{\varphi }}+3\frac{\dot{a}}{a}\dot{\bar{\varphi }}-\frac{\dot{\mathcal{N}}}{\mathcal{N}}\dot{\bar{\varphi }}+\mathcal{N}^2 V_{,\varphi }-\frac{3f f_{,\varphi }}{a^2 }\dot{A}^2 =0, \\
 && \frac{af^2}{\mathcal{N}} \dot{A} = {\rm const.} \equiv c M_{\rm pl}. \label{eq:1stint}
 \end{eqnarray}
The first integral (\ref{eq:1stint}) will be used to eliminate $\dot{A}$. In order to figure out 
the conditions for inflation, let $t$ be the proper time coordinate by choosing 
$\mathcal{N} =1$ and define the Hubble expansion rate $H = \dot{a}/a$. For the spatial slice to 
undergo accelerated expansion and for it to last more than one Hubble time, it is required that
the parameters
 \begin{eqnarray}
 \epsilon _{H} &=& -\frac{\dot{H}}{H^2}  \ = \ \frac{\dot{\bar{\varphi }}^2}{2M^2_{\rm pl} H^2} +\frac{c^2}{a^4f^2 H^2}, \\
 \eta _{H} &=& \frac{\dot{\epsilon }_{H}}{H\epsilon _{H}} \ = \ 2\left( \epsilon _{H}+ \frac{\dot{\bar{\varphi }}^2}{2\epsilon _{H}M^2_{\rm pl}H^2} \frac{\ddot{\bar{\varphi }}}{H\dot{\bar{\varphi }}} \right) -2(1-\frac{\dot{\bar{\varphi }}^2}{2\epsilon _{H}M^2_{\rm pl}H^2} )\left( 2+\frac{\dot{f}}{Hf} \right) 
 \end{eqnarray}
be much less than unity. These two parameters characterize the evolution of the space-time: 
$\epsilon _H \ll 1$ guarantees accelerated expansion $\ddot{a}/a >0$ and constancy of $H$ 
over a few Hubble times and $\eta _H \ll 1$ ensures that this regime lasts for at least tens of 
e-foldings. 
Another important element is the balance between the scalar kinetic energy and the amplitude of
the gauge fields. It proves to be convenient to use the following set of parameters;
 \begin{eqnarray}
 \epsilon _{\varphi } &=& \frac{\dot{\bar{\varphi }}^2}{2M^2_{\rm pl}H^2} \ < \ \epsilon _{H} ,\\
 \eta _{\varphi } &=& \frac{\dot{\epsilon }_{\varphi }}{H\epsilon _{\varphi }} \ = \ 2\left( \epsilon _{H}+\frac{\ddot{\bar{\varphi }}}{H\dot{\bar{\varphi }}}\right) .
 \end{eqnarray}
$\epsilon _{\varphi }$ represents the kinetic energy of inflaton, which is already much smaller 
than unity by the conditions above. Its significance resides in the fact that it controls the balance 
between the kinetic energy of the inflaton and the gauge fields through
 \begin{equation}
\frac{f^2 \dot{A}^2}{a^2 M_{\rm pl}^2 H^2} =  \frac{c^2 M_{\rm pl}^2 }{a^4 f^2 H^2 } = \epsilon _{H} -\epsilon _{\varphi } <\epsilon _H .
 \end{equation}
In connection to this balance, it later proves to be useful to define another parameter
\begin{equation}
\mathcal{I} = \sqrt{\frac{\epsilon _{H}-\epsilon _{\varphi }}{\epsilon _{\varphi }}} , \label{eq:scri}
\end{equation}
measuring the ratio of kinetic energy between the background gauge fields and the inflaton. 
Note that there is no a priori constraint on $\mathcal{I}$ as long as $\epsilon _{\varphi } < \epsilon _H \ll 1$ 
is satisfied. Although it would not be necessary to assume $\eta _{\varphi } \ll 1$, 
a large $\eta _{\varphi }$ means rapidly varying $\epsilon _{\varphi }$ and in turn a rapid 
exchange of the energy between the gauge fields and the inflaton, which would require 
a dedicated condition to be met by $V$ and $f$. In order to avoid unnecessary complications, 
we do assume the smallness of $\eta _{\varphi }$, which leads to the control over the gradients
of $V$ and $f$ through
\begin{equation}
\frac{V_{, \varphi }}{M_{\rm pl}H^2} = \frac{1}{\sqrt{2\epsilon _{\varphi }}}\left( -6\epsilon _{H}+3\epsilon _{H}^2 -\frac{3}{2}\epsilon _{H}\eta _{H}-\epsilon _{H}\epsilon _{\varphi }+\frac{1}{2}\epsilon _{\varphi }\eta _{\varphi } \right)
\end{equation}
and
\begin{equation}
\frac{M_{\rm pl}f_{\varphi }}{f} = \frac{1}{\sqrt{2\epsilon _{\varphi }}}\left( -2 +\epsilon _{H}-\frac{1}{2}\frac{\epsilon _{H}\eta _{H}-\epsilon _{\varphi }\eta _{\varphi }}{\epsilon _{H}-\epsilon _{\varphi }}\right) .\label{eq:gaugeenergy}
\end{equation}
The slower $\varphi $ rolls down the potential, the greater is the amplitude of the gauge fields and 
the steeper is the slope of the gauge-kinetic function.

\section{The Second Order Action and Scalar-Vector-Tensor Decomposition}
Now let us perturb the background and write down the second order action. Our convention for the perturbative variables are given as follows:
\begin{description}
\item[ Metric ]
\begin{equation}
N = \mathcal{N}(1+\phi ), \ \ \ \ \ N_i = \mathcal{N}a \beta _i , \ \ \ \ \ g_{ij} = a^{2}(\delta _{ij} +2\gamma _{ij} ) .
\end{equation} 
\item[ Inflaton ]
\begin{equation}
\varphi = \bar{\varphi } + \pi .
\end{equation}
\item[ Gauge fields ]
\begin{equation}
A^{(m)}_0 = \sigma ^m , \ \ \ \ \ A^{(m)}_i = A \delta ^m _{\ i} + \chi ^m _{\ i}.
\end{equation}
\end{description}
Note that the definition of $\phi $ is unconventional. The derivation for the gravity and 
scalar actions can be seen in any standard literature (e.g. ref. \cite{Feldman}). To derive the gauge-field Lagrangian, it is convenient to define the perturbed electric fields
\begin{equation}
X^m _{\ i} = \dot{\chi }^m_{\ i} - \sigma ^m _{\ ,i}.
\end{equation}
Substituting
\begin{equation}
E^{(m)}_{i} = \dot{A}\delta ^m_{\ i} + X^m_{\ i} ,  \ \ \ \ \ F^{(m)}_{\ \ ij} = -2 \chi ^m_{\ [i,j]} 
\end{equation}
and
\begin{equation}
 E^{(m)}_{i} + F^{(m)}_{\ \ ij}N^j = \dot{A}\delta ^m_{\ i} + X^m_{\ i} -2\frac{\mathcal{N}}{a}\chi ^m_{\ [i,j]}\beta _j + {\rm higher \ order \ terms},
 \end{equation}
into the action (\ref{eq:maxwell}), one obtains
\begin{eqnarray}
\mathcal{L}^{(2)}_M &=& \frac{af^2}{2\mathcal{N}}X^m_{\ i}X^m_{\ i}-\frac{af^2}{\mathcal{N}}\dot{A}\left( 2\gamma _{ij}X^{i}_{\ j}-\gamma X^m _{\ m} \right) +\frac{a\dot{A}}{\mathcal{N}}\left( 2f f_{,\varphi }\pi - f^2\phi \right) X^{m}_{\ m} \nonumber \\
&& -\frac{\mathcal{N}f^2}{a} \chi ^m _{\ [i,j]}\chi ^m _{\ [i,j]} -2f^2\dot{A}\chi ^i _{\ [i,j]}\beta _j +\frac{3}{2}\frac{af^2}{\mathcal{N}}\dot{A}^2 \phi ^2 -\frac{a\dot{A}^2}{2\mathcal{N}}\phi \left( f^2\gamma +6f f_{,\varphi }\pi \right) \\
&& -\frac{af^2}{2\mathcal{N}}\dot{A}^2 \left( \frac{1}{2}\gamma ^2 -\gamma _{ij}\gamma _{ij} \right) + \frac{af f_{,\varphi }}{\mathcal{N}}\dot{A}^2 \pi \gamma + \frac{3}{4}\frac{a(f^2)_{,\varphi \varphi }}{\mathcal{N}}\dot{A}^2 \pi ^2 \nonumber
\end{eqnarray}
where $\gamma = \gamma _{ii}$. Combining this with the gravity and scalar Lagrangians, 
using the background equations and performing integration by parts, we derive the general
quadratic Lagrangian as
\begin{eqnarray}
\mathcal{L}^{(2)} &=& \frac{a^3 M_{\rm pl}^2 }{2\mathcal{N}}\left( \dot{\gamma }_{ij}\dot{\gamma }_{ij}-\dot{\gamma }^2 \right) +\frac{\mathcal{N}aM_{\rm pl}^2}{2}\left( 2\gamma _{ij,j}\gamma _{ik,k}-\gamma _{ij,k}\gamma _{ij,k}+2\gamma \gamma _{ij.ij}-\gamma \gamma _{,ii} \right) \nonumber \\
&& + \frac{af^2}{2\mathcal{N}}\dot{A}^2 \left( 2\gamma _{ij}\gamma _{ij}-\gamma ^2 \right) + \frac{a^3}{2\mathcal{N}}\dot{\pi }^2 -\frac{\mathcal{N}a}{2}\pi _{,i}\pi _{,i} -\frac{1}{2}\left( \mathcal{N}a^3 V_{,\varphi \varphi }-\frac{3a(f f_{,\varphi })_{,\varphi }}{\mathcal{N}}\dot{A}^2 \right) \pi ^2 \nonumber \\
&&  +\frac{af^2}{2\mathcal{N}}X^i_{\ j}X^i_{\ j}-\frac{\mathcal{N}f^2}{a}\chi ^k_{\ [i,j]} \chi ^k_{\ [i,j]} -2\frac{af}{\mathcal{N}}\dot{A}\gamma _{ij}X^i_{\ j} +\frac{2af f_{,\varphi } }{\mathcal{N}}\dot{A}\pi X^m _{\ m} \nonumber \\
&& + \frac{\mathcal{N}aM_{\rm pl}^2 }{2}\left( \beta _{(i,j)}\beta _{(i,j)}-\beta _{i,i}^2 \right) +a^2 M_{\rm pl}^2 \beta _i \left( \dot{\gamma }_{ij,j}-\dot{\gamma }_{,i}+2\frac{\dot{a}}{a}\phi _{,i} \right) \nonumber \\
&& -\mathcal{N}a^3 V\phi ^2 +\gamma \left[ \frac{a^3 \dot{\bar{\varphi }}}{\mathcal{N}}\dot{\pi }-\left( \mathcal{N}a^3 V_{,\varphi }-\frac{aff_{,\varphi }}{\mathcal{N}}\dot{A}^2 \right) \pi +\frac{af^2}{\mathcal{N}}\dot{A}X^m_{\ m} \right] \nonumber \\
&& -a^2\beta _i \left( \dot{\bar{\varphi }}\pi _{,i}+2\frac{f^2}{a}\dot{A}\chi ^k_{\ [k,i]} \right)  +M_{\rm pl}^2 \phi \left[ \frac{2a^2\dot{a}}{\mathcal{N}}\dot{\gamma } +\mathcal{N}a \left( \gamma _{ij,ij}-\gamma _{,ii} \right) \right] \\
&& +\phi \left[ \frac{af^2}{\mathcal{N}}\dot{A}^2 \gamma  -\frac{a^3 \dot{\bar{\varphi }}}{\mathcal{N}}\dot{\pi } -\left( \mathcal{N}a^3 V_{,\varphi }+\frac{3a\dot{A}^2 f f_{,\varphi }}{\mathcal{N}} \right) \pi -\frac{af^2}{\mathcal{N}}\dot{A}X^m _{\ m} \right] .\nonumber
\end{eqnarray}
Although the upper indices originated from the label attached 
to each copy of gauge field, some of them are now contracted with spatial indices. This happened 
since the background respects the spatial $O(3)$ symmetry and is also invariant under linear 
mixing of the three vector fields by $ O(3)$ "rotation," namely $A^{(m)}_{\ \ \ i} \propto \delta ^m_{\ i}$, 
which resulted in transferring spatial indices to the ones denoting species. Consequently, 
$\chi ^i_{\ j}$ (and $X^i_{\ j}$) looks as if it were a $3\times 3$ spatial tensor even though 
it represents a triplet of vectors. This suggests that one can formally extend the 
scalar-vector-tensor decomposition of the perturbed quantities as was first carried out 
in the ref. \cite{Iran}. We use the following symbols;
\begin{eqnarray}
\beta _ i &=& B_{,i}-S_i ,\\
\gamma _{ij} &=& -\psi \delta _{ij}+E_{,ij}+F_{(i,j)}+\frac{1}{2}h_{ij}, \\
\chi ^i_{\ j} &=& \alpha \delta _{ij} + \theta _{,ij}+ \epsilon _{ijk}\tau _{,k}+\kappa _{(i,j)}+\epsilon _{ijk}\lambda _k + \omega _{ij} , \\
\sigma ^i &=& \mu _{,i}+\nu _i .
\end{eqnarray}
As usual, $S_i, F_i, \kappa _i, \lambda _i$ and $\nu _i$ are divergence-free vectors and 
$h_{ij}$ and $\omega _{ij}$ are traceless transverse tensors and $\epsilon _{ijk}$ is the 
Levi-Civita symbol in three dimensions. Now all the indices are downstairs, which signals 
we are going to ignore the distinction between spatial and component indices. It can be easily
seen that the Lagrangian indeed splits into three pieces each of which contains only scalars, 
vectors and tensors respectively, up to a surface term, and one can deal with each mode separately as far as linear perturbations are concerned.

Not all of the quantities are dynamical because of the gauge freedom of general relativity and
the $U(1)$ gauge symmetry. Let us first consider an infinitesimal space-time diffeomorphism
\begin{equation}
t \rightarrow t + \eta , \ \ \ \ \ x^i \rightarrow x^i + \xi _{,i} + \xi _i .
\end{equation}
It induces the following transformations:
\begin{eqnarray}
&& \mu \rightarrow \mu + A\dot{\xi }, \ \ \ \ \ \nu _{i} \rightarrow \nu _{i} +A\dot{\xi }_i , \nonumber \\
&& \alpha \rightarrow \alpha -\dot{A}\eta , \ \ \ \ \ \theta \rightarrow \theta +A\xi \\
&& \kappa _i \rightarrow \kappa _i + A\xi _i , \ \ \ \ \ \lambda _i \rightarrow \lambda _i +\frac{A}{2}\epsilon _{ijk}\xi _{j,k} . \nonumber
\end{eqnarray}
Note that $\tau $ as well as $\omega _{ij}$ are gauge invariant. One is also allowed to perform the infinitesimal $U(1)$ transformations
\begin{equation}
A^{(m)}_{\ \ \ \mu } \rightarrow A^{(m)}_{\ \ \ \mu } + \partial _{\mu }\rho ^{(m)} 
\end{equation}
where $\rho ^{(m)}$ are arbitrary scalar functions. It is again convenient to decompose them as
\begin{equation}
\rho ^{(m)} = \rho _{,m} + \rho _m 
\end{equation}
and the transformation laws become
\begin{eqnarray}
&& \mu \rightarrow \mu + \dot{\rho }, \ \ \ \ \ \nu _i \rightarrow \nu _i +\dot{\rho }_i , \nonumber \\
&& \theta  \rightarrow \theta + \rho, \ \ \ \ \ \kappa _i \rightarrow \kappa _i + \rho _i , \\
&& \lambda _i \rightarrow \lambda _i +\frac{1}{2}\epsilon _{ijk}\rho _{j,k} . \nonumber
\end{eqnarray}

It is instructive to count the numbers of degrees of freedom. There are four scalars from the metric, 
four from the gauge fields and another for the inflaton. Two from the metric ($\phi , B$) and one from the 
gauge fields ($\mu $) are non-dynamical. We have three arbitrary functions for gauge transformations,
which results in three dynamical scalar degrees of freedom. For vector perturbations, there are 
two from the metric and three from the gauge-fields, among which we have one non-dynamical for each
sector ($S_i$ and $\nu _i$). Two more are redundant because of the gauge freedom and we are left with only one
dynamical vector mode. Finally, there are two gauge-invariant dynamical tensor degrees of 
freedom.

\section{Scalar Modes and the Curvature Power Spectrum}
In this section, we consider the scalar perturbation; $\phi, \psi, B, E, \pi , \alpha , \theta , \tau $ and $\mu $ and compute the curvature power spectrum. We fix the gauge by setting
\begin{equation}
\psi = E =\dot{\theta } +\mu =0 ,
\end{equation}
which means the constant-time hypersurfaces are flat and $\pi , \alpha $ and $\tau $ are the
three dynamical variables. The convenient choice of time-coordinate is the 
conformal time $\eta $ ($\mathcal{N}=a$) and differentiation with respect to it is denoted by 
primes. We set $M_{\rm pl}=1$ in this section and the next to save the space. 

\subsection{The Lagrangian and the Curvature of the Uniform-density Hypersurfaces}
The second order Lagrangian is written as
\begin{eqnarray}
\mathcal{L}_{S}^{(2)} &=& \frac{a^2}{2}\left( \pi ^{\prime 2} -\pi _{.i}\pi _{,i} \right) -\frac{1}{2}\left( a^4 V_{,\varphi \varphi }-3\left(f f_{,\varphi }\right) _{,\varphi }A^{\prime 2} \right) \pi ^2 +6f f_{,\varphi }A^{\prime }\pi \alpha ^{\prime } \nonumber \\
&& + f^2 \left( \frac{3}{2}\alpha ^{\prime 2} -\alpha _{,i}\alpha _{,i}+\tau ^{\prime }_{,i}\tau ^{\prime }_{,i}-\tau _{,ii}\tau _{,jj}\right) -a^2 B \left( 2\frac{a^{\prime }}{a}\phi _{,ii}-\bar{\varphi }^{\prime }\pi _{,ii}-2\frac{f^2}{a^2}A^{\prime }\alpha _{,ii} \right) \\
&& -a^4 V \phi ^2 - \phi \left[ a^2 \bar{\varphi }^{\prime }\pi ^{\prime }+\left(a^4 V_{,\varphi }+ 3f f_{,\varphi }A^{\prime 2} \right) \pi + 3f^2 A^{\prime }\alpha ^{\prime } \right] .\nonumber 
\end{eqnarray}
Varying $B$ gives
\begin{equation}
\phi = \frac{\bar{\varphi }^{\prime }}{2\mathcal{H}}\pi +\frac{f^2}{\mathcal{H}a^2}A^{\prime }\alpha 
\end{equation}
where we defined
\begin{equation}
\mathcal{H} = \frac{a^{\prime }}{a} .
\end{equation}
Its substitution into the action eliminates the non-dynamical variables $\phi $ and $B$ and leaves 
three scalar degrees of freedom. Using the canonically normalized variables
\begin{eqnarray}
 \hat{\pi } &= & a \pi, \nonumber \\
 \hat{\alpha } &= & \sqrt{3}f\alpha ,\\
 \hat{\tau } &= & \sqrt{2}f\tau, \nonumber
 \end{eqnarray}
and performing several integrations by parts, it yields
\begin{eqnarray}
\mathcal{L}^{(2)}_S &=& \frac{1}{2}\bigg( \hat{\pi }^{\prime 2} -\left( \nabla \hat{\pi } \right) ^2 -m^2_{\pi } \hat{\pi }^2 \bigg)+\frac{1}{2}\left( \hat{\alpha }^{\prime 2}-\frac{2}{3}\left( \nabla \hat{\alpha } \right) ^2-m^2_{\alpha }\hat{\alpha }^2 \right)  \nonumber \\
&& -\sqrt{\frac{2\epsilon _{\varphi }(\epsilon _{H} -\epsilon _{\varphi })}{3} }\mathcal{H} \hat{\pi }^{\prime }\hat{\alpha } +\sqrt{3(\epsilon _H -\epsilon _{\varphi })}\left( 2\frac{f_{,\varphi }}{f}-\sqrt{\frac{\epsilon _{\varphi }}{2}} \right) \mathcal{H}\hat{\pi }\hat{\alpha }^{\prime } +g_{\pi \alpha } \hat{\pi }\hat{\alpha } \\
&& + \frac{1}{2}\left( (\nabla\hat{\tau })^{\prime 2} - (\nabla ^2 \hat{\tau })^2 +\frac{f^{\prime \prime }}{f}(\nabla \hat{\tau })^2 \right) , \nonumber
\end{eqnarray}
where 
\begin{eqnarray}
m^2_{\pi } &=&  -(2+3\epsilon _{\varphi }-\epsilon _{H}-\epsilon _{\varphi }\epsilon _{H}+\eta _{\varphi }\epsilon _{\varphi } )\mathcal{H}^2 \nonumber \\
&& + a^2 \left( V_{\varphi \varphi }+\sqrt{2\epsilon _{\varphi }}V_{\varphi }+\epsilon _{\varphi }V \right) -3(\epsilon _{H}-\epsilon _{\varphi })\left( \frac{f_{\varphi \varphi }}{f}+\frac{f^2_{\varphi }}{f^2}-\sqrt{2\epsilon _{\varphi }}\frac{f_{\varphi }}{f}\right) \mathcal{H}^2 , \nonumber \\
m^2_{\alpha } &=&  - \frac{f^{\prime \prime }}{f}+(\epsilon _{H}-\epsilon _{\varphi })\left( 3-\epsilon _{H} \right)\mathcal{H}^2  -\frac{2}{3}\left( \epsilon _{H}-\epsilon _{\varphi } \right) a^2 V, \\
g_{\pi \alpha } &=& \sqrt{\frac{2\epsilon _{\varphi }(\epsilon _{H}-\epsilon _{\varphi })}{3}} \left( \mathcal{H}^2 - a^2 V\right) -\sqrt{3(\epsilon _{H}-\epsilon _{\varphi })}\left( 2\frac{f_{,\varphi }}{f}-\sqrt{\frac{\epsilon _{\varphi }}{2}} \right) \frac{f^{\prime }}{f} \mathcal{H}  \nonumber \\
&& -\sqrt{\frac{\epsilon _{H}-\epsilon _{\varphi }}{3}}\left( a^2 V_{,\varphi }+3\frac{f_{,\varphi }}{f} \left( \epsilon _{H}-\epsilon _{\varphi } \right) \mathcal{H}^2 \right) .\nonumber
\end{eqnarray}
We notice that $\hat{\tau }$ is decoupled from the other degrees of freedom since it is a pseudo-scalar representing 
the magnetic fields, which are zero in the background and hence gauge invariant. As its evolution 
is trivial and only affects the curvature perturbation at a non-linear order, we ignore this contribution.
It should be noted, however, that it may well play an important role in the three-point or 
higher order correlation functions. On the other hand, $\hat{\alpha }$ makes a physical contribution 
to the scalar curvature at linear order directly (see the equation (\ref{eq:density})) and through 
the interaction with $\hat{\pi }$, even though its origin is the vectorial gauge fields. It is not surprising since 
this $\hat{\alpha }$-mode represents the modulation of the background value of the gauge triad 
which already affects the background expansion. $\hat{\alpha }$ is a superposition of three 
mutually orthogonal vector modes, which also explains why its propagation velocity is reduced 
from the speed of light to $c_s^2 =  2/3$.

Our aim is to compute the curvature perturbation in the uniform density gauge, which is given by
\begin{equation}
- \zeta = \psi + \mathcal{H}\frac{\delta \rho }{\bar{\rho }^{\prime }}
\end{equation}
in the linear order according to Malik and Wands \cite{Malik}. This quantity is not conserved during the inflationary 
period in the present model, even on the super-horizon scales because of the multi-field 
interaction and we will calculate its spectrum at the end of inflation quantum mechanically 
by tree-level perturbative expansion of the in-in formalism. Although it might be
 modified even after inflation by 
the entropy perturbations potentially generated by the gauge fields, we will not discuss it 
as the details depend on reheating. The definition of the total energy perturbation 
$\delta \rho $ here is that of Kodama and Sasaki \cite{Sasaki}. At linear order, it is just the 0-0 component 
of the energy-momentum tensor and we have
\begin{eqnarray}
\delta \rho &=& - \delta T^{0 }_{\ 0} \label{eq:density} \\ 
&=& a^{-2}\bar{\varphi }^{\prime } \pi ^{\prime }+\left[ -\frac{\bar{\varphi }^{\prime 3}}{2\mathcal{H}a^2}+\frac{3c^2}{a^4 f^2}\left( \frac{f_{\varphi }}{f}-\frac{\bar{\varphi }^{\prime }}{2\mathcal{H}} \right) + V_{\varphi } \right] \pi + \frac{3c}{a^4} \alpha ^{\prime } -\frac{c}{\mathcal{H}a^4} \left( \bar{\varphi }^{\prime 2} +\frac{3c^2}{a^2 f^2} \right) \alpha  \nonumber \\
&=&  a^{-3}\sqrt{2\epsilon _{\varphi }} \mathcal{H} \hat{\pi} ^{\prime }+a^{-3}\left[ -\sqrt{2\epsilon _{\varphi }}(1-\epsilon _{\varphi })\mathcal{H}^2 +3(\epsilon _{H}-\epsilon _{\varphi } )\left( \frac{f_{\varphi }}{f}-\sqrt{\frac{\epsilon _{\varphi }}{2}} \right) \mathcal{H}^2 + a^2 V_{\varphi } \right] \hat{\pi } \nonumber \\
&& + a^{-3}\sqrt{3(\epsilon _{H}-\epsilon _{\varphi })} \mathcal{H} \hat{\alpha }^{\prime }+a^{-3}\sqrt{\frac{\epsilon _{H}-\epsilon _{\varphi }}{3}}\left( -\frac{3f^{\prime }}{f} \mathcal{H} + (3\epsilon _{H}-\epsilon _{\varphi } )\mathcal{H}^2 \right) \hat{\alpha }. \nonumber
\end{eqnarray}
So far, we haven't used the smallness of the four parameters, 
$\epsilon _{H}, \epsilon _{\varphi }, \eta _H $, and $\eta _{\varphi }$. Keeping the leading order 
terms in these parameters and assuming 
 \begin{equation}
 \dot{\eta }_{H} = O(1)H\eta _{H } , \ \ \ \ \ \dot{\eta }_{\varphi } = O(1)H\eta _{\varphi }, \label{eq:eta}
 \end{equation}
which is reasonable, the curvature perturbation is expressed as
\begin{equation}
\zeta = \frac{1}{3\mathcal{H}a\epsilon _{H}} \sqrt{\frac{\epsilon _{\varphi }}{2}} \left[ \hat{\pi }^{\prime }+\frac{4+6\mathcal{I}^2 }{\eta } \hat{\pi }+ \sqrt{\frac{3}{2}}\mathcal{I} \left( \hat{\alpha }^{\prime }-\frac{2}{\eta }\hat{\alpha }\right) \right] , \label{eq:zeta}
\end{equation}
whose evolution is governed by the Lagrangian
\begin{equation}
\mathcal{L} = \frac{1}{2}\hat{\pi }^{\prime 2} -\frac{1}{2} (\nabla \hat{\pi })^2 +\frac{1+ 6\mathcal{I}^2 }{\eta ^2 }\hat{\pi }^2   + \frac{1}{2}\hat{\alpha }^{\prime 2} -\frac{1}{3}(\nabla \hat{\alpha })^2 +\frac{1}{\eta ^2 } \hat{\alpha }^2 + \frac{2\sqrt{6}\mathcal{I}}{\eta }  \hat{\pi }\hat{\alpha }^{\prime } -\frac{4\sqrt{6}\mathcal{I}}{\eta ^2}\hat{\pi }\hat{\alpha }.
\end{equation}
The steepness of the gauge-kinetic function compensates the suppressions coming from 
the small background density of the inflaton and the gauge fields as well as the flatness of the 
inflaton potential, hence results in the apperance of the parameter $\mathcal{I}$ defined 
in (\ref{eq:scri}), which is potentially of order unity.

\subsection{Calculation of the Power Spectrum}
In order to calculate the two-point function, we treat the terms involving $\mathcal{I}$ as small perturbations.
Then, $\hat{\pi }$ and $\hat{\alpha }$ are free particles in the de-Sitter background at the leading 
order. We impose the Bunch-Davies vacuum condition and write them as
\begin{eqnarray}
\hat{\pi }(\eta , \bf{x} ) &=& \int \frac{d^3 k}{(2\pi )^3}\left( u_k (\eta ) a_{\bf{k}} e^{i \bf{k}\cdot \bf{x}} + u^{\ast }_{k} (\eta ) a^{\dagger }_{\bf{k}}e^{-i\bf{k}\cdot \bf{x}} \right) , \\
\hat{\alpha }(\eta , \bf{x} ) &=& \int \frac{d^3 k}{(2\pi )^3}\left( \tilde{u}_k (\eta ) b_{\bf{k}} e^{i \bf{k}\cdot \bf{x}} + \tilde{u}^{\ast }_{k} (\eta ) b^{\dagger }_{\bf{k}}e^{-i\bf{k}\cdot \bf{x}} \right) .
\end{eqnarray}
In the above expressions and all the following, the field symbols denote the corresponding 
quantities in the interaction picture. The vacuum mode functions are given by
\begin{eqnarray}
u_{k}(\eta ) &=& \frac{1}{\sqrt{2k}}\left( 1-\frac{i}{k\eta } \right) e^{-ik\eta } , \\
\tilde{u}_k(\eta ) &=& \frac{1}{\sqrt{2c_s k}}\left( 1 -\frac{i}{c_s k\eta } \right) e^{-ic_s k\eta } 
\end{eqnarray}
with $c_s^2 =2/3$ being the propagation speed of $\hat{\alpha }$, 
and the creation and annihilation operators satisfy
\begin{equation}
[ a_{\bf{p}} , a^{\dagger }_{\bf{q}} ] = [ b_{\bf{p}} , b^{\dagger }_{\bf{q}} ] = (2\pi )^3 \delta (\textbf{p}-\textbf{q} ) , \ \ \ \ \ [ a_{\bf{p}}, b_{\bf{q}} ] = 0, \ \ \ \ \ {\rm etc}.
\end{equation}
The interaction Hamiltonian is given by
\begin{equation}
H_{I}(\eta ) = \int d^3 x \left( -\frac{6\mathcal{I}^2}{\eta ^2 }\hat{\pi }^2 + \frac{2\sqrt{6}\mathcal{I}}{\eta }\hat{\pi }^{\prime }\hat{\alpha }+\frac{2\sqrt{6}\mathcal{I}}{\eta ^2}\hat{\pi }\hat{\alpha } \right) . \label{eq:hamiltonian}
\end{equation}
Note that we performed integration by parts for simplifying the calculations below. In interpreting 
the expression, normal ordering is understood in order to eliminate vacuum bubbles. 
In terms of the de-Sitter vacuum state defined by
\begin{equation}
a_{\bf{k}} |0 \rangle = b_{\bf{k}}|0\rangle  = 0 ,
\end{equation}
the in-state is written as
\begin{equation}
| {\rm in}\rangle = T\exp \left( -i \int ^{\eta }_{-\infty (1-i\epsilon )} d\tilde{\eta } H_I (\tilde{\eta }) \right) |0\rangle 
\end{equation}
where the symbol $T \exp $ denotes the time-ordered exponential. While the lower limit of the time
 integral should physically be the beginning of inflation, it does not affect the following calculations 
 as long as it is far enough in the past. The quadratic variation of 
the curvature perturbation during inflation is defined by
\begin{eqnarray}
(2\pi )^3 \delta (\textbf{k}+\textbf{p}) \langle {\rm in} |\zeta _k^2 (\eta ) |{\rm in} \rangle &=& \int d^3 x \ e^{i\bf{k}\cdot \bf{x}}\langle {\rm in }|\zeta (\eta , \textbf{x}) \zeta (\eta , 0) |{\rm in} \rangle \nonumber \\
&=& \int \frac{d^3 p}{(2\pi )^3}\langle {\rm in}| (\zeta _{\bf{k}}+\zeta ^{\dagger }_{\bf{k}})(\zeta _{\bf{p}}+\zeta ^{\dagger }_{-\bf{p}}) | {\rm in}\rangle 
\end{eqnarray}
where we used
\begin{equation}
\zeta (\eta , \textbf{x}) = \int \frac{d^3 k}{(2\pi )^3}\left( \zeta _{\bf{k}}(\eta )e^{-i\bf{k}\cdot \bf{x}}+\zeta _{\bf{k}}(\eta )^{\dagger }e^{-\bf{k}\cdot \bf{x}} \right) .
\end{equation}
Comparing this with (\ref{eq:zeta}), we need to compute
\begin{eqnarray}
\langle {\rm in}| \left( g_k (\eta )a_{\bf{k}}+g^{\ast }_k (\eta )a^{\dagger }_{-\bf{k}} \right) \left( g_p (\eta )a_{\bf{p}}+g^{\ast }_p (\eta )a^{\dagger }_{-\bf{p}} \right) |{\rm in} \rangle, \label{eq:pipi} \\
\langle {\rm in}| \left( j_k (\eta )b_{\bf{k}}+j^{\ast }_k (\eta )b^{\dagger }_{-\bf{k}} \right) \left( j_p (\eta )b_{\bf{p}}+j^{\ast }_p (\eta )b^{\dagger }_{-\bf{p}} \right) |{\rm in} \rangle, \label{eq:alal}
\end{eqnarray}
and
\begin{equation}
\langle {\rm in}| \left( g_k (\eta )a_{\bf{k}}+g^{\ast }_k (\eta )a^{\dagger }_{-\bf{k}} \right) \left( j_p (\eta )b_{\bf{p}}+j^{\ast }_p (\eta )b^{\dagger }_{-\bf{p}} \right) |{\rm in} \rangle, \label{eq:pial}
\end{equation}
where $g_k (\eta )$ and $j_k (\eta )$ are linear combinations of $u_k(\eta )$, $\tilde{u}_k(\eta )^{\prime }$ and their derivatives given by
\begin{eqnarray}
g_k (\eta ) &=& u^{\prime }_k (\eta )+\frac{4+6 \mathcal{I}^2 }{\eta } u_k (\eta ) \\
 j_k (\eta ) &=&  \tilde{u}^{\prime }_k (\eta ) -\frac{2}{\eta }\tilde{u}_k (\eta ) .
\end{eqnarray}

Let us start from (\ref{eq:pipi}). Expanding the time ordered exponential up to second order in $\mathcal{I}$, 
we need to keep the following contributions:
\begin{eqnarray}
&& |g_k (\eta )|^2 (2\pi )^3 \delta ( \bf{k}+ \bf{p} ) \\
&& + i \langle 0|  \int ^{\eta } d\eta _1 H_{I}(\eta _1 ) \left( g_k (\eta ) a_{\bf{k}}+g^{\ast }_k (\eta )a^{\dagger }_{-\bf{k}} \right) g^{\ast }_p (\eta ) a^{\dagger }_{-\bf{p}} |0\rangle \label{eq:1stA} \\
&& - i\langle 0| g_k (\eta )a_{\bf{k}}\left( g_p (\eta ) a_{\bf{p}}+g^{\ast }_p (\eta )a^{\dagger }_{-\bf{p}} \right)\int ^{\eta } d\eta _1 H_{I}(\eta _1) |0\rangle \label{eq:1stB} \\
&& - \langle 0|  \int ^{\eta } d\eta _1 \int ^{\eta _1} d\eta _2 H_I (\eta _2) H_{I}(\eta _1 ) \left( g_k (\eta ) a_{\bf{k}}+g^{\ast }_k (\eta )a^{\dagger }_{-\bf{k}} \right) g^{\ast }_p (\eta ) a^{\dagger }_{-\bf{p}} |0\rangle \label{eq:2ndA} \\
&& - \langle 0| g_k (\eta )a_{\bf{k}}\left( g_p (\eta ) a_{\bf{p}}+g^{\ast }_p (\eta )a^{\dagger }_{-\bf{p}} \right)\int ^{\eta } d\eta _1 \int ^{\eta _1} d\eta _2 H_{I}(\eta _1) H_I (\eta _2) |0\rangle \label{eq:2ndB} \\
&& + \langle 0| \int ^{\eta }d\eta _1 H_I (\eta _1 ) \left( g_k a_{\bf{k}} + g_k^{\ast }a^{\dagger }_{-\bf{k}} \right) \left( g_p a_{\bf{p}}+g^{\ast }_p a^{\dagger }_{-\bf{p}} \right) \int ^{\eta }d\eta _2 H_I (\eta _2) | 0\rangle . \label{eq:2ndC}
\end{eqnarray}
The first term in (\ref{eq:hamiltonian}), which can be written in momentum space as
\begin{equation}
-\frac{6\mathcal{I}^2}{\eta ^2}\int d^3 x \hat{\pi }^2 = -\frac{6\mathcal{I}^2}{\eta ^2}\int \frac{d^3 k}{(2 \pi )^3}\left( u_k^2 a_{\bf{k}}a_{-\bf{k}} +2 |u_k|^2 a^{\dagger }_{\bf{k}}a_{\bf{k}} +u^{\ast 2}_{k} a^{\dagger }_{\bf{k}}a^{\dagger }_{-\bf{k}} \right) , \label{eq:mass}
\end{equation}
contributes to the second and third lines only at the order $\mathcal{I}^2$. 
This modified mass effect reduces to 
\begin{eqnarray}
(\ref{eq:1stA}) +(\ref{eq:1stB}) &=& 24 \mathcal{I}^2 (2\pi )^3 \delta (\textbf{k}+\textbf{p}) \int ^{\eta }d\eta _1 \frac{1}{\eta _1^2} \mathfrak{Im}\left( u_k (\eta _1 )^2 g^{\ast }_k (\eta )^2 \right) ,
\end{eqnarray}
where $\mathfrak{Im}$ means taking the imaginary part. The integration can be readily carried 
out by using the cosine- and sine-integrals that are defined by
\begin{eqnarray}
{\rm Ci}(x) &=& \int ^x_{\infty } dt \frac{\cos t}{t} \ = \ \gamma + \ln x + O(x^2), \\
{\rm Si}(x) &=& -\int ^x_{\infty }dt \frac{\sin t}{t} \ = \ \frac{\pi }{2} + O(x) .
\end{eqnarray}
It is convenient to introduce $x= k\eta $ and the rational functions
\begin{eqnarray}
P(x) &=& \left( 1+\frac{3(1+2\mathcal{I}^2)}{x^2 } \right) ^2 -\frac{9(1+2\mathcal{I}^2 )^2}{x^2}, \\
Q(x) &=& \frac{3(1+2\mathcal{I}^2 )}{x} \left( 1+\frac{3(1+2\mathcal{I}^2)}{x^2} \right) ,
\end{eqnarray}
by which $g_k^{\ast 2} $ is expressed as
\begin{equation}
g^{\ast }_k (\eta )^2 = \frac{k}{2} \left( -P(x) +2iQ(x) \right) e^{2ix} .
\end{equation} 
We obtain 
\begin{equation}
24 \mathcal{I}^2  \int ^{\eta }d\eta _1 \frac{1}{\eta _1^2} \mathfrak{Im}\left( u_k (\eta _1 )^2 g^{\ast }_k (\eta )^2 \right) = 4\mathcal{I}^2 k \mathcal{M} 
\end{equation}
where the amplitude $\mathcal{M}$ is given by
\begin{equation}
\mathcal{M} =  -\left( P\cos 2x +2 Q \sin 2x \right) {\rm Ci}(-2x) + \left( P\sin 2x - 2Q \cos 2x \right) {\rm Si}(-2x) -\frac{1}{x^2}P +\left( \frac{1}{x^3} +\frac{1}{x} \right) Q .
\end{equation}
The cross terms in the Hamiltonian (\ref{eq:hamiltonian}) in Fourier space become
\begin{equation}
\frac{2\sqrt{6}\mathcal{I}}{\eta }\int d^3 x \left( \hat{\pi }^{\prime }\hat{\alpha }+\frac{1}{\eta } \hat{\pi } \hat{\alpha }\right) = \frac{2\sqrt{6}\mathcal{I}}{\eta }\int \frac{d^3 k}{(2\pi )^3}\left( v_k \tilde{u}_k a_{\bf{k}}b_{-\bf{k}}+ v_k \tilde{u}_k ^{\ast }b^{\dagger }_{\bf{k}}a_{\bf{k}}+ v_k^{\ast }\tilde{u}_k a^{\dagger }_{\bf{k}}b_{\bf{k}} + v_k^{\ast }\tilde{u}^{\ast }_k a^{\dagger }_{\bf{k}}b^{\dagger }_{-\bf{k}} \right) , \label{eq:cross}
\end{equation}
where
\begin{eqnarray}
v_k (\eta ) &=& u^{\prime }_k (\eta )+\frac{1}{\eta }u_k (\eta ) .
\end{eqnarray}
They do not affect (\ref{eq:1stA}) and (\ref{eq:1stB}) at tree-level but are responsible for the remaining 
terms. After eliminating the contributions corresponding to disconnected diagrams, they yield
 \begin{eqnarray}
 (\ref{eq:2ndA}) +(\ref{eq:2ndB}) +(\ref{eq:2ndC}) &=& -48 \mathcal{I}^2 (2\pi )^3  \delta (\textbf{k}+\textbf{p}) \nonumber \\
 &\times &\int ^{\eta }d\eta _1 \int ^{\eta _1}d\eta _2 \frac{1}{\eta _1 \eta _2}\left(  v_k (\eta _2 )u_k (\eta _2) v_k (\eta _1 ) u^{\ast }_k (\eta _1) g_k^{\ast }(\eta )^2 +({\rm c.c.}) \right) \\
 && + 48 \mathcal{I}^2 (2\pi )^3 \delta (\textbf{k}+ \textbf{p} ) \int ^{\eta }\frac{d\eta _1}{\eta _1} \int ^{\eta }\frac{d\eta _2}{\eta _2} v_k (\eta _1 ) u_k (\eta _1 ) v^{\ast }_k (\eta _2) u^{\ast }_k (\eta _2) | g_k (\eta ) |^2 \nonumber .
 \end{eqnarray}
The integrals are not as bad as they look once the relations
\begin{equation}
v_k (\eta _1 ) \tilde{u}_k^{\ast } (\eta _1 )  = \frac{-i}{2\sqrt{c_s}}\left( 1+\frac{i}{c_s k\eta _1}\right) e^{-i(1-c_s )k\eta _1} ,
 \end{equation}
and
 \begin{equation}
v_k (\eta _2 ) \tilde{u}_k (\eta _2 ) = \frac{-i}{2\sqrt{c_s }}\left( 1-\frac{i}{c_s k\eta _2} \right) e^{-i(1+c_s )k\eta _2} 
 \end{equation}
are taken into account. The result is
\begin{equation}
(\ref{eq:2ndA}) + (\ref{eq:2ndB}) + (\ref{eq:2ndC}) = 6\mathcal{I}^2 k \mathcal{A}  (2\pi )^3 \delta (\textbf{k}+\textbf{p} )
\end{equation}
with the amplitude
\begin{eqnarray}
&& \mathcal{A}  = \frac{4(3+\mathcal{I})^2}{c_s^3 x^3 }\left[ \frac{1}{2x} -{\rm Ci}(-(1+c_s)x)\sin (1+c_s)x + {\rm Si}(-(1+c_s)x)\cos (1+c_s)x \right] \nonumber \\
&& +\frac{8}{c_s^2 } Q(x) \left[ \frac{1}{2x} -{\rm Ci}(-2x)\sin 2x + {\rm Si}(-2x)\cos 2x \right] -\frac{4}{c_s^2}P(x) \left[ {\rm Ci}(-2x) \cos 2x + {\rm Si}(-2x)\sin 2x \right] \nonumber \\
&& -\frac{4}{c_s^3 x}Q(x) \left[   {\rm Ci}(-(1+c_s )x) \cos (1+c_s)x  + {\rm Si}(-(1+c_s)x) \sin (1+c_s)x\right] \nonumber \\
&&  -\frac{2}{c_s^3} \int ^x \frac{dy}{y}\int ^y \frac{dz}{z} \left[ P(x)\sin (2x-y-z) -2Q(x)\cos (2x-y-z) \right] \sin c_s (y-z) \\
&& + \frac{1}{c_s^3} \left( P(x) \sin 2x -2Q(x) \cos 2x \right) \left[ {\rm Ci}(-(1-c_s)x) {\rm Si}(-(1+c_s)x) - {\rm Si}(-(1-c_s)x) {\rm Ci}(-(1+c_s)x) \right] \nonumber \\
&& + \frac{1}{c_s^3}\left( P(x) \cos 2x +2Q(x)\sin 2x \right) \left[ {\rm Ci}(-(1-c_s)x){\rm Ci}(-(1+c_s)x)-{\rm Si}(-(1-c_s)x){\rm Si}(-(1+c_s)x) \right] \nonumber \\
&& + \frac{1}{c_s^3}\left( P(x)+\frac{2(3+\mathcal{I}^2)^2}{x^2} \right) \left( {\rm Ci}(-(1+c_s)x)^2 + {\rm Si}(-(1+c_s)x)^2 \right)  .\nonumber
\end{eqnarray}
The $\alpha $-$\alpha $ correlation (\ref{eq:alal}) is needed only at the leading order and given by
\begin{equation}
(\ref{eq:alal}) = |j_k(\eta )\mathcal{|}^2 (2\pi )^3 \delta (\textbf{k}+\textbf{p}) .
\end{equation}
Finally, the cross correlation (\ref{eq:pial}) is calculated as
\begin{eqnarray}
(\ref{eq:pial}) &=& i \langle 0| \int ^{\eta }d\eta _1 H_I (\eta _1)\left( g_k a_{\bf{k}} +g^{\ast }_k a^{\dagger }_{-\bf{k}}\right) j^{\ast }_p b^{\dagger }_{-\bf{p}} | 0\rangle + ({\rm h.c.}) \nonumber \\
 && + i \langle 0 | \int ^{\eta }d\eta _1 H_I (\eta _1) \left( j_k b_{\bf{k}}+ j^{\ast }_k b^{\dagger }_{-\bf{k}} \right) g^{\ast }_p a^{\dagger }_{-\bf{p}} | 0\rangle + ({\rm h.c.}) \nonumber \\
 &=& 4\sqrt{6}\mathcal{I} i \int ^{\eta }\frac{d\eta _1}{\eta _1} \left[ v^{\ast }_k (\eta _1 ) \tilde{u}^{\ast }_k(\eta _1)g_k (\eta ) j_k (\eta ) - ({\rm c.c.}) \right] \ \equiv \ 2\sqrt{6}\mathcal{I}k \mathcal{B} 
 \end{eqnarray}
 with $\mathcal{B}$ given by
  \begin{eqnarray}
 \mathcal{B} &=& \frac{3+\mathcal{I}^2}{c_s x^2}\left( 1-\frac{3}{c_s^2 x^2} \right) -\frac{3}{c_s^2 x} \left( 1+\frac{3+\mathcal{I}^2}{x^2} \right) \\
 &-& \frac{1}{c_s}\left\{ \frac{3(3+\mathcal{I}^2)}{c_s x^2}+\left( 1+\frac{3+\mathcal{I}^2}{x^2}\right) \left( 1-\frac{3}{c_s^2 x^2} \right) \right\} \nonumber \\
 && \times  \left( {\rm Ci}(-(1+c_s)x) \cos (1+c_s)x -{\rm Si}(-(1+c_s)x)\sin (1+c_s)x \right)  \nonumber \\
 &+&  \frac{1}{c_s} \left\{ \frac{3+\mathcal{I}^2}{x} \left( 1-\frac{3}{c_s^2 x^2} \right) -\frac{3}{c_s x} \left( 1+\frac{3+\mathcal{I}^2}{x^2} \right) \right\} \nonumber \\
 && \times \left( {\rm Si}(-(1+c_s)x)\cos (1+c_s)x + {\rm Ci}(-(1+c_s)x)\sin 1+c_s)x  \right) .\nonumber 
 \end{eqnarray}
Combining all the pieces, 
the curvature power spectrum is given by
 \begin{eqnarray}
 \langle {\rm in}| \zeta _k (\eta )^2 | {\rm in}\rangle = \frac{H^2 \epsilon _{\varphi }}{18\epsilon _{H}^2} \eta ^4 \left( |g_k(\eta )|^2 +4\mathcal{I}^2 k \mathcal{M}+ \frac{3\mathcal{I}^2 }{2} | j_k (\eta )|^2 + 6\mathcal{I}^2 k \mathcal{A} +12\mathcal{I}^2k \mathcal{B} \right) ,
 \end{eqnarray}
 where we used $a=-1/H\eta $. In the standard calculation for single-field models, one evaluates 
 this quantity at the horizon crossing since $\zeta $ is conserved, even at non-linear order. Here 
 in contrast, $\zeta $ still evolves beyond the horizon scale and therefore we are going to evaluate 
 its value at the end of inflation $\eta = \eta _f$. The scales of cosmological interest should be 
well outside the horizon at the end of inflation, 
which means $-k\eta _f \ll 1$. In the corresponding limit $x \rightarrow 0$, the power spectrum reduces to
\begin{equation}
 \langle {\rm in}| \zeta _k (\eta _f )^2 | {\rm in}\rangle \rightarrow \frac{\epsilon _{\varphi }}{\epsilon _H^2}\frac{H^2}{4k^3} \left( 1 + 18\sqrt{6}\mathcal{I}^2 \left( \ln |k\eta _f | \right) ^2 \right) . \label{result1}
\end{equation}
 The disadvantage in this strategy is to introduce 
extra errors by neglecting time-variation of $H$ and the slow-roll parameters over a number of 
Hubble times. We assume that these corrections are subdominant since they are suppressed by
$\epsilon _H$ and the other small parameters. In fact, the leading correction coming from varying $H$  
is proportional to $\epsilon _{H} \ln |k\eta |$, which can be safely discarded compared to $\mathcal{I}^2 (\ln |k\eta |)^2$. It is difficult to estimate the effect of varying $\mathcal{I}$
as it is already a correction from the non-adiabatic evolution of $\zeta $. It is expected to
be proportional to
\begin{equation}
\frac{\dot{\mathcal{I}}}{H\mathcal{I}} = \frac{\epsilon_H}{2}\frac{\eta _H -\eta _{\varphi }}{\epsilon _H -\epsilon _{\varphi }} 
\end{equation}
multiplied by several powers of $\ln |k\eta |$. Thus the approximation might break down when
the e-folding number is too large.

\section{Primordial Gravitational Waves and Phenomenological Implications}

\subsection{Tensor Power Spectrum}
Let us now turn to the tensor perturbations
\begin{equation}
\gamma _{ij} = \frac{1}{2}h_{ij}, \ \ \ \ \ X^i_{\ j} = \omega _{ij}
\end{equation}
where both $h_{ij}$ and $\omega _{ij}$ are trace- and diergence-free. 
Our quadratic action is
\begin{eqnarray}
\mathcal{L}^{(2)}_{T} = \frac{a^2}{8}\left( h^{\prime }_{ij} h^{\prime }_{ij}- h_{ij,k}h_{ij,k} \right) +\frac{c^2}{4f^2} h_{ij}h_{ij} +\frac{f^2}{2}\left( \omega ^{\prime }_{ij}\omega ^{\prime }_{ij}-\omega _{ij,k}\omega _{ij,k} \right) -c\ h_{ij}\omega ^{\prime }_{ij}  .
\end{eqnarray}
 Normalizing the variables by
 \begin{equation}
 \hat{h}_{ij} = \frac{1}{2}ah_{ij}, \ \ \ \ \ \hat{\omega }_{ij} =f\omega _{ij}, 
 \end{equation}
 it becomes
 \begin{eqnarray}
 \mathcal{L}^{(2)}_{T} &=& \frac{1}{2}\left[ \hat{h}^{\prime }_{ij}\hat{h}^{\prime }_{ij}-\hat{h}_{ij,k}\hat{h}_{ij,k}+\left( \frac{a^{\prime \prime }}{a}+\frac{2c^2 }{f^2 a^2 } \right) \hat{h}_{ij}\hat{h}_{ij} \right] \\
 &&+ \frac{1}{2}\left( \hat{\omega }^{\prime }_{ij}\hat{\omega }^{\prime }_{ij}-\hat{\omega }_{ij,k}
\hat{\omega }_{ij,k}+\frac{f^{\prime \prime }}{f}\hat{\omega }_{ij}\hat{\omega }_{ij} \right) -\frac{2c}{af}\left( \hat{\omega }^{\prime }_{ij}-\frac{f^{\prime }}{f}\hat{\omega }_{ij} \right) \hat{h}_{ij} .
 \end{eqnarray}
 They are decomposed according to their polarization as
 \begin{equation}
 \hat{h}_{ij}(\eta ,\textbf{x}) = \sum _{s=1,2}\epsilon _{ij}^s \hat{h}^s (\eta ,\textbf{x}), \ \ \ \ \ \epsilon _{ij}^s \epsilon _{ij}^{s^{\prime }} =\delta _{s s^{\prime }} ,
 \end{equation}
 and a similar decomposition applies to $\hat{\omega }_{ij}$ as well. Applying Fourier transform as before,
\begin{equation}
\hat{h}^s (\eta , \textbf{x}) = \int \frac{d^3k}{(2\pi )^3}\left( \hat{h}^s_{\bf{k}}(\eta ) e^{i\bf{k}\cdot \bf{x}} + \hat{h}^{s\dagger }_{\bf{k}}(\eta ) e^{-i\bf{k}\cdot \bf{x}} \right) ,
\end{equation} 
what one wants to compute here is the amplitude of the gravitational waves
\begin{equation}
(2\pi )^3 \delta (\textbf{k}+\textbf{p}) \langle {\rm in}| h_k(\eta )^2 |{\rm in}\rangle =\frac{4}{a^2}\sum _{s=1,2}\int \frac{d^3 p}{(2\pi )^3} \langle {\rm in}| \left( \hat{h}_{\bf{k}}^s (\eta ) +\hat{h}_{-\bf{k}}^{s \dagger }(\eta ) \right) \left( \hat{h}_{\bf{p}}^s (\eta ) +\hat{h}_{-\bf{p}}^{s \dagger }(\eta ) \right) |{\rm in}\rangle .
\end{equation}
For this purpose, we only have to keep the leading order terms in $\hat{\omega }_{ij}$ and therefore
\begin{equation}
\frac{f^{\prime }}{f} \sim -2\mathcal{H}, \ \ \ \ \ \frac{f^{\prime \prime }}{f} \sim 2\mathcal{H}^2 .
\end{equation}
After discarding higher order corrections, the problem becomes a quantum field theory for the 
Lagrangian
 \begin{eqnarray}
 \mathcal{L} &=& \frac{1}{2}\sum _{s=1,2}\left[ \left( (\hat{h}^{s} ) ^{\prime }\right) ^{ 2}- (\nabla \hat{h}^s )^2 +\frac{2-\epsilon _{\varphi }}{\eta ^2}(\hat{h}^s)^{2}  + \left( (\hat{\omega }^s ) ^{\prime }\right) ^2-(\nabla \hat{\omega }^s)^2 +\frac{2}{\eta ^2}(\hat{\omega }^s )^2\right] \\
 &&  +\frac{2}{\eta }\sqrt{\epsilon _{H}-\epsilon _{\varphi }}\sum _{s=1,2}\left( \left( \hat{\omega }^s \right) ^{\prime }-\frac{2}{\eta }\hat{\omega }^s \right) \hat{h}^s . \nonumber
 \end{eqnarray}
One notices that the two polarization decouple from each other and the Lagrangian is of 
the same form as the scalar modes with different coupling constants and twice many fields. 
The interaction Hamiltonian for the each polarization mode is given by
\begin{equation}
H_I^s = \frac{\epsilon _{\varphi }}{2\eta ^2} \int d^3 x (\hat{h}^s )^2 + \frac{2\sqrt{\epsilon _H -\epsilon _{\varphi }}}{\eta } \int d^3x \left( (\hat{h}^s )^{\prime }\hat{\omega }^s + \frac{1}{\eta }\hat{h}^s \hat{\omega }^s \right) .
\end{equation}
Hence, if we focus on the corrections up to first order in $\epsilon _H$ and $\epsilon _{\varphi }$, 
we only have to repeat the calculations in the previous section with replacements
\begin{equation}
\frac{\epsilon _{\varphi }}{2\eta ^2 }\int \frac{d^3 k}{(2 \pi )^3}\left( u_k^2 a^s_{\bf{k}}a^s_{-\bf{k}} +2 |u_k|^2 a^{s \dagger }_{\bf{k}}a^s_{\bf{k}} +u^{\ast 2}_{k} a^{s \dagger }_{\bf{k}}a^{s \dagger }_{-\bf{k}} \right) \ \ \ {\rm for} \ \ \ (\ref{eq:mass})
\end{equation}
and
\begin{equation}
\frac{2\sqrt{\epsilon _H -\epsilon _{\varphi}}}{\eta }\int \frac{d^3 k}{(2\pi )^3}\left( v_k u_k a^s_{\bf{k}}b^s_{-\bf{k}}+ v_k u_k ^{\ast }b^{s \dagger }_{\bf{k}}a^s_{\bf{k}}+ v_k^{\ast }u_k a^{s\dagger }_{\bf{k}}b^s_{\bf{k}} + v_k^{\ast }u^{\ast }_k a^{s\dagger }_{\bf{k}}b^{s\dagger }_{-\bf{k}} \right) \ \ \ {\rm for} \ \ \ (\ref{eq:cross}),
\end{equation}
and multiply the result by a factor of two to add up the polarizations. 
We also substitute $u_k(\eta ) $ for $g_k(\eta )$, which simplifies the calculations. In the end, one obtains
\begin{equation}
\langle {\rm in} | h_k (\eta  )^2 |{\rm in}\rangle = \frac{4H^2 \eta ^2}{k} \left( 1 +\frac{1}{x^2} -\frac{2}{3}\epsilon _{\varphi }\tilde{\mathcal{M}} +2(\epsilon _H -\epsilon _{\varphi })\tilde{\mathcal{A}} \right) 
\end{equation}
with the amplitudes given by
\begin{eqnarray}
\tilde{\mathcal{M}} &=& \frac{2}{x^2} -\frac{1-x^2}{x^2} \left( {\rm Ci}(-2x) \cos 2x - {\rm Si}(-2x) \sin 2x \right) \\
&& -\frac{2}{x} \left( {\rm Ci}(-2x)\sin 2x + {\rm Si}(-2x) \cos 2x \right) , \nonumber \\
\tilde{\mathcal{A}} &=& \frac{2}{x^2} -\frac{2(1-x^2)}{x^3} \sin  2x +\frac{4}{x}\cos 2x + \frac{1+x^2}{x^2}\left( {\rm Ci}(-2x)^2 + {\rm Si}(-2x)^2 \right) \\
&& + \left( \frac{1-x^2}{x^2} \ln | x| -\frac{8}{x^2}+4 \right)\left( {\rm Ci}(-2x)\cos 2x + {\rm Si}(-2x)\sin 2x \right) \nonumber  \\
&& + \frac{2}{x}\left( \ln |x| -6 \right) \left( {\rm Ci}(-2x)\sin 2x -{\rm Si}(-2x)\cos 2x \right) \nonumber \\
&& +2\int ^x \frac{dy}{y}\int ^y \frac{dz}{z}\left[ \frac{2}{x}\cos (2x-y-z)-\frac{1-x^2}{x^2}\sin (2x-y-z) \right] \sin (y-z) \nonumber . 
\end{eqnarray}
In the limit $x \rightarrow 0$, it becomes
\begin{equation}
\langle {\rm in} | h_k (\eta _f )^2 |{\rm in}\rangle \rightarrow \frac{4H^2}{k^3} \left[ 1 + 4(\epsilon _H -\epsilon _{\varphi }) \left( \ln |k\eta _f | \right) ^2 \right] . \label{result2}
\end{equation}

\subsection{Vector Mode}
Before discussing the observational implications, we shall briefly look at the vector perturbation 
 \begin{eqnarray}
&& \beta _i = -S_i, \ \ \ \ \ \gamma _{ij} = F_{(i,j)} , \\
&& \sigma ^i = \nu _i , \ \ \ \ \ \chi ^i_{\ j} = \kappa _{(i,j)} +\epsilon _{ijk}\lambda _k ,
\end{eqnarray}
for completeness.
As before, we take the flat slicing $F_{i}=0$.  The Lagrangian is given by
 \begin{eqnarray}
 \mathcal{L}^{(2)} &=&  +\frac{f^2}{4}\kappa ^{\prime }_{i,j}\kappa ^{\prime }_{i,j} - \frac{f^2}{8}\kappa _{i,jk}\kappa _{i,jk}+f^2 \lambda ^{\prime }_k \lambda ^{\prime }_k -\frac{f^2}{2} \lambda _{i,j}\lambda _{i,j} \nonumber \\
 && -\frac{f^2}{2} \epsilon _{ijk}\lambda _{k,l}\kappa _{j,il}+\frac{a^2}{4}S_{i,j}S_{i,j}-a^2 S_i \left( \frac{f^2}{a}A^{\prime }\kappa _{i,jj}+\frac{2f^2}{a}A^{\prime }\epsilon _{ijk}\lambda _{j,k} \right) \\
 && +\frac{f^2}{2}\nu _{i,j}\nu_{i,j} + \nu _{i,j}\left( f^2 \kappa ^{\prime }_{(i,j)}+f^2 \epsilon _{ijk}\lambda ^{\prime }_k  \right) . \nonumber
 \end{eqnarray}
 Varying $S_i$ yields
 \begin{equation}
 \nabla ^2 S_i = -\frac{2f^2}{a}A^{\prime } \left( \nabla ^2 \kappa _i + 2{\rm curl} \lambda _i \right) .
 \end{equation}
 Similarly, $\nu _i$ is non-dynamical and solved as
 \begin{equation}
 \nabla ^2 \nu _i = -\frac{1}{2}\nabla ^2 \kappa ^{\prime }_i + {\rm curl}\lambda _i^{\prime } .
 \end{equation}
 Perhaps there are two promising choices of the gauge for the vector fields; $\lambda _i =0$ and $\kappa _i =0$. 
 The action becomes 
 \begin{equation}
 \mathcal{L} = \frac{f^2}{4} \left( \nabla \kappa ^{\prime }_i \right) \cdot \left( \nabla \kappa ^{\prime }_i \right) -\frac{f^2}{8}\left( \nabla ^2 \kappa _i \right) \left( \nabla ^2 \kappa _i \right) 
 \end{equation}
 for the former and 
 \begin{equation}
 \mathcal{L} = f^2 \lambda ^{\prime }_k \lambda ^{\prime }_k -\frac{f^2}{2}\left( \nabla \lambda _i \right) \cdot \left( \nabla \lambda _i \right) 
 \end{equation}
 for the latter. In either way, we have a free massless vector field with the propagation speed $c_s = 1/\sqrt{2}$. 
 
\subsection{Phenomenological Consequences}
If we assume instantaneous reheating whereby all the energy of the inflaton and gauge fields
is dumped into a single relativistic fluid, the scalar curvature $\zeta $ and the gravitational wave 
$h_{ij}$ are conserved until re-entering inside the Hubble horizon and are observable through
the CMB. The vector perturbation generated by the quantum fluctuation will quickly decay away and not be observed. For the scalar mode, the relevant parameter is the spectral 
tilt $n_S -1$. From (\ref{result1}), this model predicts
\begin{equation}
n_S -1 = \frac{d}{d \ln k} \ln \left( k^3 \langle \zeta _k^2 \rangle \right) = \frac{36\sqrt{6}\mathcal{I}^2 \ln |k\eta _f |}{1+18\sqrt{6} \mathcal{I}^2 \left( \ln |k\eta _f | \right) ^2 } ,
\end{equation}
which is negative for $|k\eta _f | \ll 1$, whence the spectrum is red. There is no contribution from
the time-variation of $H, \epsilon _{H,\varphi }$ and $\mathcal{I}$ since the spectrum was evaluated at the end of 
inflation, not the time each mode crossed the horizon. It can be seen that
\begin{equation}
| n_S -1| \leq  \min \left( -36\sqrt{6}\mathcal{I}^2 \ln |k\eta _f | , -\frac{2}{\ln |k\eta _f |} \right) ,
\end{equation}
which means the spectral tilt doesn't necessarily impose a stringent constraint on the value 
of $\mathcal{I}$. We note that $\ln |k\eta _f |$ is the e-folding number counted from horizon exit of 
the mode with wavenumber $k$ until the end of inflation, whose value is model-dependent.
If, for instance, we take $\ln |k\eta _f | \sim -50$, it yields
\begin{equation}
|n_S -1| \sim \frac{1}{25}
\end{equation}
which is nicely consistent with WMAP 7-year \cite{Komatsu}, even with $\mathcal{I} \sim 1$. The running of $n_S$ can
also be computed as
\begin{equation}
\frac{d}{d\ln k} \left( n_S -1 \right)  = \frac{n_S -1}{\ln |k \eta _f |} - (n_S -1)^2 ,
\end{equation}
which is safely small as long as $n_S -1$ is small. Similarly, the tilt of the tensor spectrum is 
given as
\begin{equation}
n_T = \frac{d}{d\ln k } \ln \left( k^3 \langle h_k^2 \rangle \right) = \frac{8(\epsilon _H -\epsilon _{\varphi } )\ln |k\eta _f |}{1+ 4(\epsilon _H -\epsilon _{\varphi } ) \left( \ln |k\eta _f | \right) ^2 } .
\end{equation}
Thus, the spectrum of gravitational wave is also red. However, it is much closer to scale 
invariance than that of scalar if $\mathcal{I} \ll 1$.

As can be seen from (\ref{result2}), the amplitude of the tensor mode itself no longer provides 
the unambiguous information about the energy scale of inflation as it receives a potentially 
significant correction. The tensor-to-scalar ratio $r$ can be used to determine $\epsilon _H$ 
and $\epsilon _{\varphi }$. Recalling $\mathcal{I}^2 = (\epsilon _H -\epsilon _{\varphi } )/\epsilon _{\varphi }$, 
it is given by
\begin{equation}
r = \frac{\langle h_k^2 \rangle }{\langle \zeta _k^2 \rangle } = 16\epsilon _H^2 \frac{1+4(\epsilon _H -\epsilon _{\varphi }) \left( \ln |k\eta _f | \right) ^2 }{\epsilon _{\varphi } + 18\sqrt{6}(\epsilon _H -\epsilon _{\varphi })\left( \ln |k\eta _f | \right)^2 } .
\end{equation}
Hence, the tensor to scalar ratio is suppressed compared to the single-field slow roll inflation.
The suppression is stronger when the e-folding number is greater and the scalar kinetic energy
is subdominant. 

In summary, we have seen that there are several different regimes that are consistent with 
the observations made so far, as far as the above three quantities are concerned. They are
classified in the following.
\begin{description}
\item[ $\mathcal{I}^2 \ll 1$ ] Recalling the formula (\ref{eq:gaugeenergy}), this occurs when the energy 
density of the gauge fields is much smaller than the scalar kinetic energy density. If the e-folding 
number experienced by the modes at CMB scales is of order hundred or so, the model predicts 
a slightly red scalar power spectrum and almost scale invariant gravitational waves. Since it means
$\epsilon _H \sim \epsilon _{\varphi }$, the tensor-to-scalar ratio is unchanged from the ordinary
single-field slow-roll inflation.  
\item[ $\mathcal{I} \sim 1$ ] Although it means the dominant contribution to the scalar power spectrum 
comes from the terms proportional to $\mathcal{I}^2$, whose origin is the interaction between the inflaton 
and the gauge fields, the model still predicts an observationally
consistent spectral tilt. On the other hand, even though the background energy density of the
gauge fields is comparable to the scalar kinetic energy, the spectrum of gravitational waves 
is not very much affected for $-\ln |k\eta _f | \sim 50$. The tensor-to-scalar ratio is
suppressed. Of course, it should be noted that the perturbative approach cannot be trusted 
and non-linear effects may significantly modify the results.
\item[ $-\ln |k\eta _f | \gg 50$]  Surprisingly, this regime is viable regardless of the value of $\mathcal{I}$. 
The spectral tilt of scalar curvature is $\sim -1/\ln |k\eta _f |$ and the tensor mode is suppressed 
by a factor of $\epsilon _H^2$ with respect to the scalar mode, which makes it practically 
undetectable. A caveat is that time-variation of the background quantities (e.g. $H$ and $\mathcal{I}$), which
was neglected in this article, might be important in this regime.
\end{description}

\section{Conclusion}
In the present article, we investigated the linear perturbation of the inflationary model with 
a triad of background gauge fields coupled to the inflaton. We characterized the accelerated
expansion by introducing four parameters $\epsilon _H , \epsilon _{\varphi }, \eta _H $ and 
$\eta _{\varphi }$ which are generalizations of the usual slow-roll parameters.  
The general second order Lagrangian was derived and irreducible mode decomposition 
was carried out according to the transformation property under the spatial rotations, 
and the internal "rotation" of the triad of gauge fields. The scalar- and tensor-mode power
spectra were computed by employing the in-in formalism on the de-Sitter background. 
We found the scalar fluctuation is characterized by the parameter $\mathcal{I}$, which is 
potentially of order unity, while its tensor counterparts are smaller by an order of $\epsilon _{\varphi }$. 
The enhancement in the scalar correction is due to the steep gauge-kinetic function. The 
observational implications were studied by looking at the spectral tilt and tensor-to-scalar 
ratio. The generic prediction is that the spectra are red with the stronger effect for the 
scalar mode and the tensor-to-scalar ratio tends to be suppressed. The magnitudes of 
tilt and the suppression depend on $\mathcal{I}$ and the e-folding number. 
The structure of the corrections is such that even accurate measurements of those 
quantities are unable to impose strong constraints on $\mathcal{I}$ or the other small parameters 
because of the involvement of the e-folding number. As it stands, both weak and strong 
background gauge fields in the unit of inflaton kinetic energy are consistent with the 
results from WMAP.

There are several implications for the preheating of auxiliary particles by the inflaton. The 
coupling needed to produce those particles back-reacts onto the scalar perturbation 
and can modify the power spectrum significantly. While gravitons can also be produced
even in this isotropic background, which could affect the determination of the energy 
scale of inflation by measuring the amplitude of tensor mode, the effect is much smaller
than that on the scalar mode, at least in the present model.

Since it appears that the model passes the observational tests at linear order, it will be 
worth looking at the higher order corrections, namely non-Gaussianity. A recent work \cite{Peloso}
suggests that the gauge-kinetic coupling can lead to a strong signal, although the analysis
was done for a vanishing background gauge field. In the above calculations, we have ignored 
the entropy modes and vector mode, which should affect the result at non-linear order. 
It is certainly interesting to carry out a thorough analysis taking into account all the modes 
involved, which should be possible for this isotropic model, and will be presented in the near future.

\begin{acknowledgements}
The initiation and the completion of this work would have been impossible had it not been for
the suggestions from Professor Jiro Soda and his contribution is appreciated. The author would 
like to thank Keiju Murata and Emanuel Malek for fruitful discussions and Sebastien Renaux-Petel 
and Hiro Funakoshi for useful comments. The author is supported by the Cambridge 
Overseas Trust.
\end{acknowledgements}

\end{document}